\documentclass[PRL,twocolumn,aps,longbibliography]{revtex4-1}
\usepackage{graphicx}
\usepackage{txfonts}
\usepackage[T1]{fontenc}
\usepackage{epstopdf}
\usepackage{verbatim}
\usepackage{color}

\makeatother
\usepackage{tikz}

\begin{document}
\title{Observation of optical vortices in momentum space}

\author{Yiwen Zhang$^{1,4}$}
\thanks{These authors contributed equally.}
\author{Ang Chen$^{1,4}$}
\thanks{These authors contributed equally.}
\author{Wenzhe Liu$^{1,4}$}
\thanks{These authors contributed equally.}
\author{Chia Wei Hsu$^{3}$}
\author{Fang Guan$^{1,4}$}
\author{Xiaohan Liu$^{1,4}$}
\author{Lei Shi$^{1,4}$}
\email{lshi@fudan.edu.cn}
\author{Ling Lu$^{2}$}
\email{linglu@iphy.ac.cn}
\author{Jian Zi$^{1,4}$}
\email{jzi@fudan.edu.cn}
\affiliation{$^{1}$ Department of Physics, Key Laboratory of Micro- and Nano-Photonic Structures (Ministry of
Education), and State Key Laboratory of Surface Physics, Fudan University, Shanghai 200433, China}
\affiliation{$^{2}$ Institute of Physics, Chinese Academy of Sciences/Beijing National Laboratory for Condensed Matter Physics, Beijing 100190, China}
\affiliation{$^{3}$ Department of Applied Physics, Yale University, New Haven, Connecticut 06520, USA}
\affiliation{$^{4}$ Collaborative Innovation Center of Advanced Microstructures, Fudan University, Shanghai 200433, China}

\begin{abstract}
\end{abstract}

\maketitle

\textbf{Vortex, the winding of a vector field in two dimensions, has its core the field singularity and its topological charge defined by the quantized winding angle of the vector field~\cite{mermin1979topological}. Vortices are one of the most fundamental topological excitations in nature, widely known in hair whorls as the winding of hair strings, in fluid dynamics as the winding of velocities, in angular-momentum beams as the winding of phase angle~\cite{nye1974dislocations,allen2003optical,yu2011light,miao2016orbital}, and in superconductors and superfluids as the winding of order parameters~\cite{kosterlitz1972long}. Nevertheless, vortices have hardly been observed other than those in the real space. Although band degeneracies, such as Dirac cones, can be viewed as momentum-space vortices in their mathematical structures, there lacks a well-defined physical observable whose winding number is an arbitrary signed integer~\cite{hsieh2009observation}. Here, we experimentally observed momentum-space vortices as the winding of far-field polarization vectors in the Brillouin zone (BZ) of periodic plasmonic structures. Using a home-made polarization-resolved momentum-space imaging spectroscopy, we completely map out the dispersion, lifetime and polarization of all radiative states at the visible wavelengths. The momentum space vortices were experimentally identified by their winding patterns in the polarization-resolved iso-frequency contours and their diverging radiative quality factors. Such polarization vortices can exist robustly on any periodic systems of vectorial fields, while they are not captured by the existing topological band theory~\cite{bansil2016colloquium} developed for scaler fields. This work opens up a promising avenue for exploring topological photonics in the momentum space~\cite{lu2014topological}, studying bound states in continuum (BICs)~\cite{hsu2016bound}, as well as for rendering and steering vector beams~\cite{yao2011orbital} and designing high-Q plasmonic resonances.}

It was recently theoretically proposed that photons in photonic crystal slabs can support
vortices with a winding far-field polarization vector in the momentum space~\cite{zhen2014topological}. Such vortices can exist at multiple momentum points across the BZ, not necessarily at the band edge~\cite{miyai2006photonics,iwahashi2011higher}.
The Bloch state at the vortex core may cease to radiate and become what is called a bound state
in the radiation continuum (BIC)~\cite{hsu2016bound,marinica2008bound,bulgakov2008bound,liu2009resonance,plotnik2011experimental,hsu2013observation,corrielli2013observation,weimann2013compact,yang2014analytical,silveirinha2014trapping,monticone2014embedded,gansch2016measurement,kodigala2017lasing,gomis2017anisotropy,xiao2017topological,bulgakov2017topological,gao2017bound}.
Moreover, these momentum-space vortices could be a general
phenomenon in vector fields, radiative or not. However, they have not been experimentally observed so far.

\begin{figure}[t]
\includegraphics[scale=1.0]{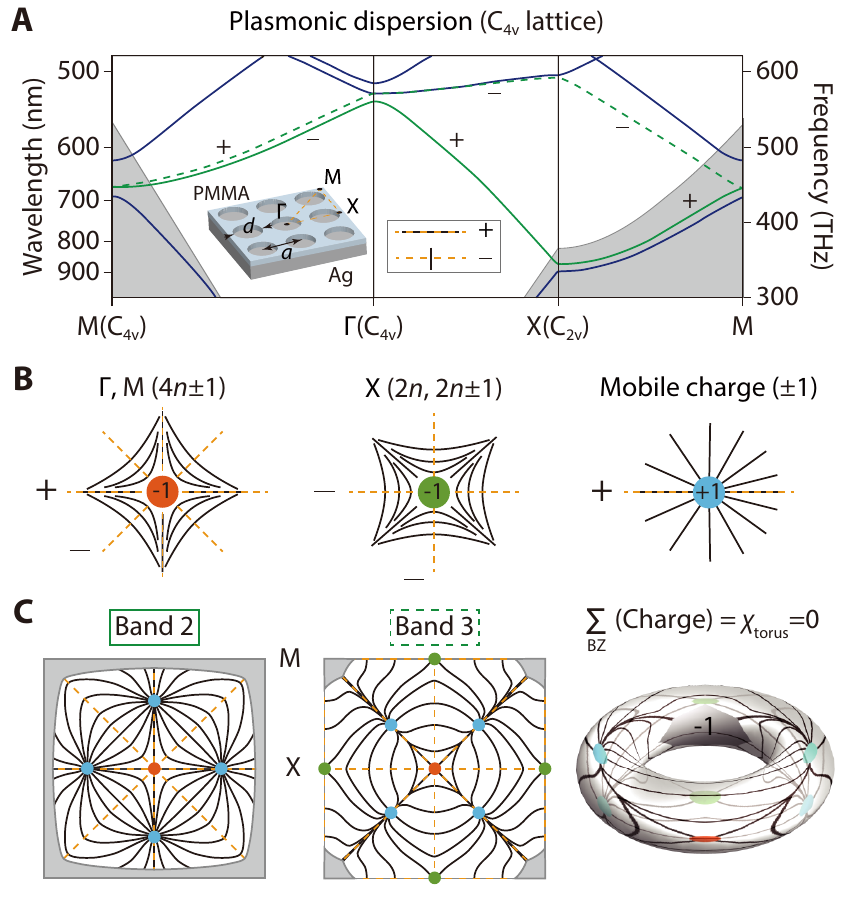}
\caption{\textbf{Theoretical analysis of topological charges in a plasmonic crystal with $C_{4v}$ symmetry.}
(\textbf{A}) Band structure of a plasmonic crystal with square lattice from FDTD simulations. A schematic view of the structure is shown in the inset.
The mirror eigenvalues~($\pm1$) of modes are labeled on the high-symmetry lines.
(\textbf{B}) Illustration of polarization vortices at high-symmetry momentum points. The $\pm$ signs indicate the mirror eigenvalues along the surrounding high-symmetry lines, and the colored circles indicate the vortices and their charges.
(\textbf{C}) Polarization streamlines computed and plotted on the whole Brillouin-zone (BZ) torus, on which the charge sum equals the Euler characteristic of a torus~($\chi=0$). The right panel shows the polarization streamlines on the 2D BZ torus of band 3.
}
\label{fig1}
\end{figure}

\begin{figure*}[t]
\includegraphics[scale=1.0]{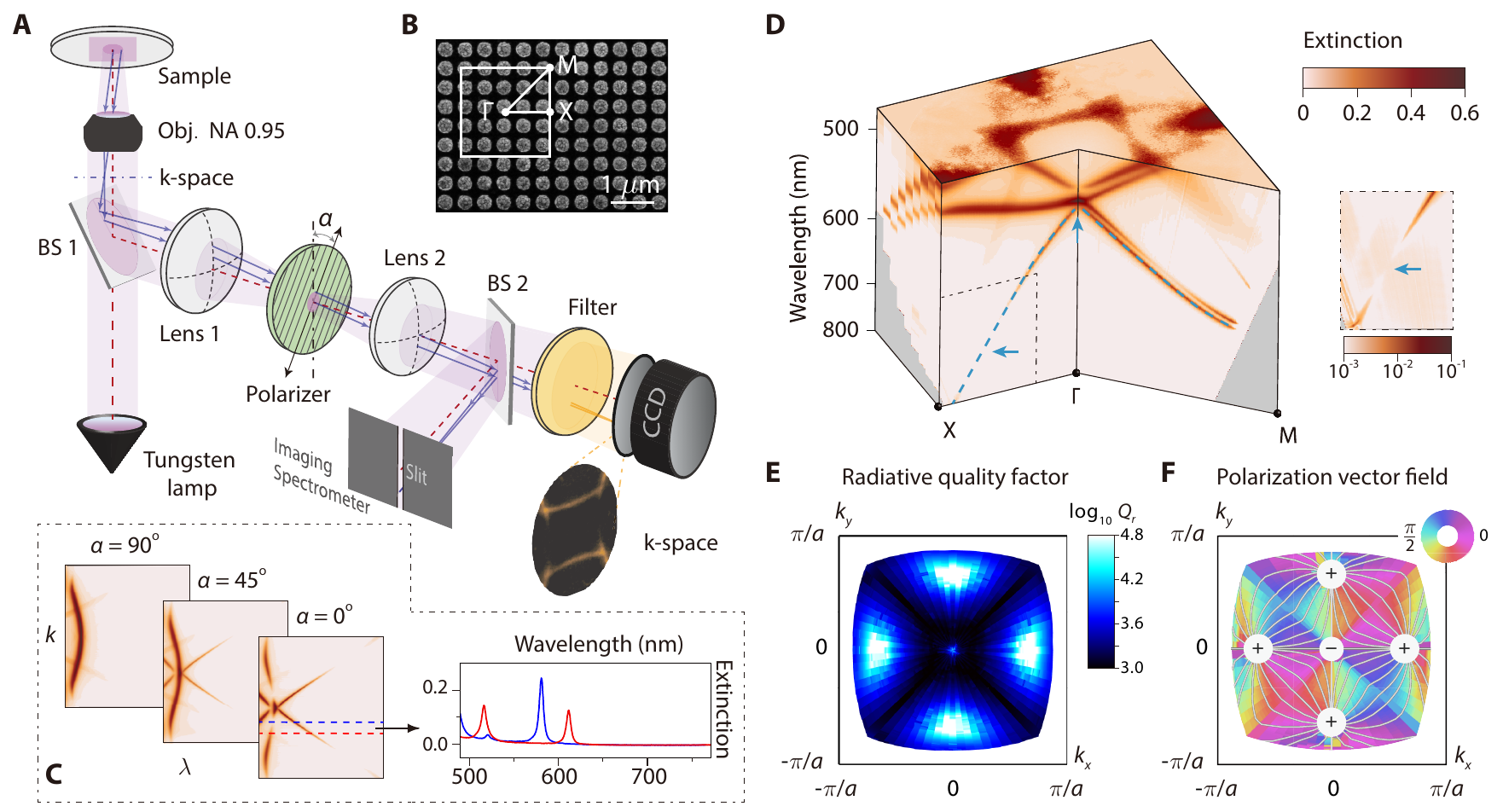}
\caption{\textbf{Experimental setup and measurement of momentum space polarization vortices in a square plasmonic lattice.} (\textbf{A}) Schematic view of the experimental setup for polarization-resolved momentum-space imaging spectroscopy. The back focal plane of the objective carries the momentum-space~(k-space) information. After lens 1 and lens 2, an imaging spectrometer measures the band dispersion, while a narrow band-pass filter and a CCD camera measure the iso-frequency contour at the selected wavelength.
A polarizer is placed on the focal plane of lens 1 to select a certain polarization. BS: beam-splitter. (\textbf{B}) SEM image of the sample. (\textbf{C}) Examples of polarization-dependent extinction spectra as a function of wavelength and wavevector for three polarizer angles of $0$, $45$ and $90$ degrees with respect to the entrance slit of the imaging spectrometer. Here, the $\Gamma$--X direction of the sample is parallel to the slit. (\textbf{D}) Measured band structure inside the whole BZ, given by the extinction spectra as a function of wavelength and wavevector under unpolarized illumination. Vortices are marked with blue arrows, with a close-up on the right. (\textbf{E}) Q factors of the lowest radiative band, as highlighted by the blue dashed lines in \textbf{D}. It corresponds to the green line~(band 2) in Fig.~\ref{fig1}A. (\textbf{F}) Angle distribution of the measured far-field polarization vectors on band 2, represented by streamlines and colors. The vortices and their topological charges are marked with $\pm$ signs.}
\label{fig2}
\end{figure*}

In this paper, we experimentally demonstrate momentum-space polarization vortices in two-dimensional (2D) plasmonic crystals at visible wavelengths. The plasmonic crystals studied here are flat metallic substrates coated with periodically corrugated thin dielectric layers fabricated using electron-beam lithography. The metallic substrate is a $200$-nm-thick silver film evaporated on a glass substrate, and the periodic dielectric layer is a $70$-nm-thick PMMA thin film~(refractive index of 1.5) with a square array of cylindrical air holes (periodicity: $400$ nm; hole diameter: $290$ nm). The metallic film is thick enough to make transmission zero. Due to the periodicity of the PMMA array, the surface plasmon polaritons~(SPPs) show well-defined band structures, and modes above the air light cone can couple into the free space and radiate~\cite{han2006transmission}.

In Fig.~\ref{fig1}A, we plotted the band structure of the plasmonic square lattice shown in the inset, calculated using the finite-difference-time-domain~(FDTD) method. Vortices of the far-field polarization vector robustly exist on each dispersion band.
We provide a general way to determine the possible vortex charges at all high-symmetry points in the BZ, regardless of the core degeneracy level, using mirror eigenvalues on the mirror-invariant momentum lines.
The $C_{4v}$ lattice has three mirror-invariant lines of $\Gamma$--M--X--$\Gamma$ in the BZ, on which the mirror eigenvalues are $\pm1$. This means the corresponding far-field polarization is either parallel or perpendicular to those momentum lines. We denote these two cases as even ($+$) and odd ($-$) representations.
By exhausting all possible combinations of representations on the high-symmetry lines surrounding each high-symmetry point, it is straightforward to see that all possible topological charges at $\Gamma$ and M points are $4n\pm1$. Here $n$ is a signed integer. Similarly, the X point has $C_{2v}$ symmetry and its possible charges are $2n\pm1$ or $2n$, detailed in the Supplementary Fig. S1.
These are consistent with the previous conclusion~\cite{zhen2014topological}.
The polarization streamlines for the lowest-order charges are illustrated in Fig.~\ref{fig1}B.
Since one cannot distinguish the electric-field direction with its opposite direction due to the time-harmonic oscillations, we represent a linear-polarization state using a line segment without arrows.
We emphasize that our discussion works for both non-degenerate and degenerate states, since the representation at the vortex core is not needed for determining its charge. When the vortex core is a non-degenerate state, its polarization singularity corresponds to a non-radiative BIC~(dark state). When the vortex core is a degenerate state, it can be either radiative (bright state) or non-radiative. In the degenerate and radiative case, the polarization is also ill-defined at the singularity. It consists of a linear combination of all polarizations and can take any polarization in the far field.

\begin{figure*}[t]
\includegraphics[scale=1.0]{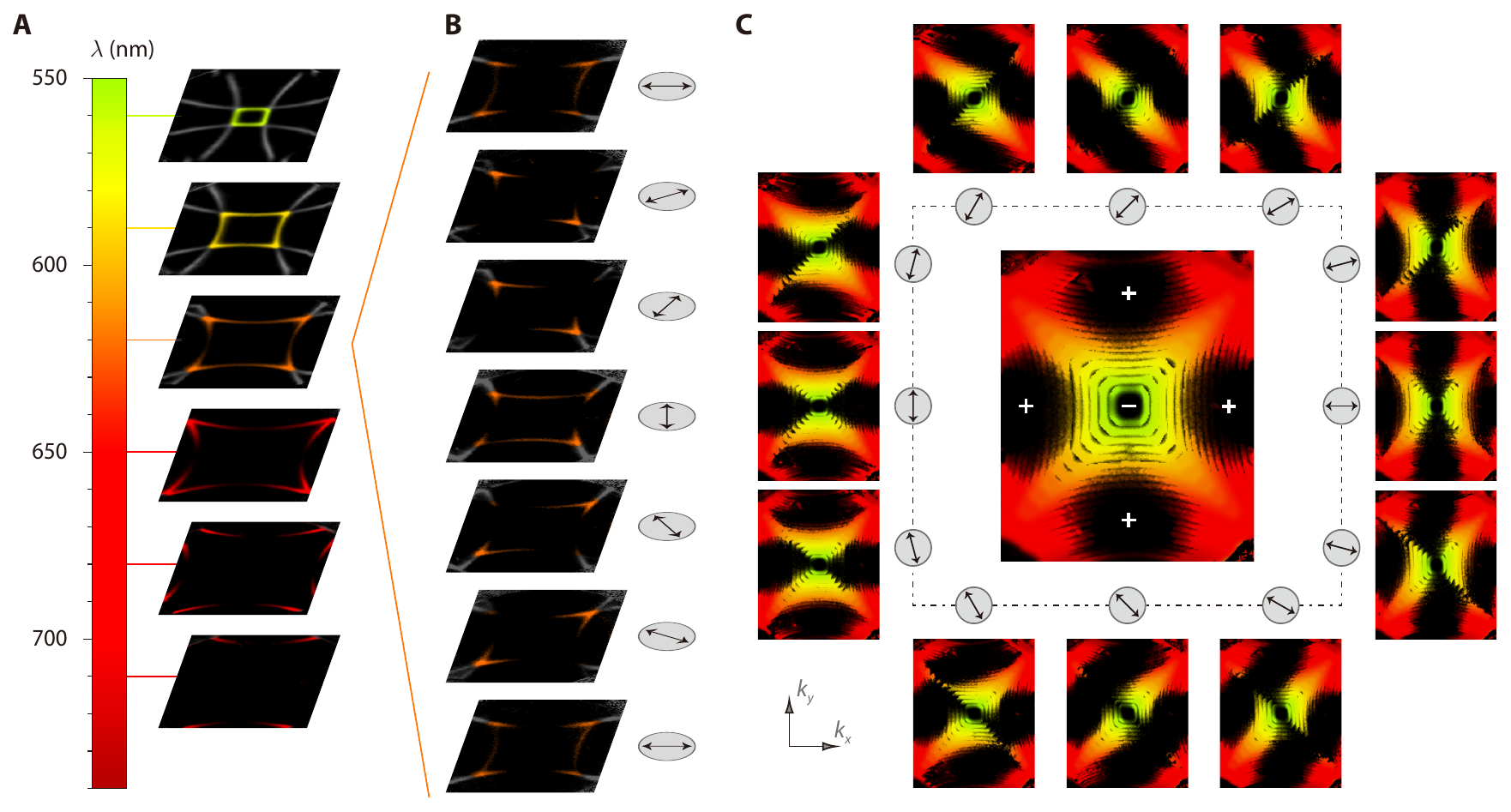}
\caption{\textbf{Direct observations of polarization vortices in the momentum space.} (\textbf{A}) Examples of measured iso-frequency contours under narrow-band illumination at wavelengths of $560$, $590$, $620$, $650$, $680$ and $710$~nm, respectively.
The colored signal is from the band of interest, and the light-gray signal is from a higher band.
(\textbf{B}) Examples of measured polarization-resolved iso-frequency contours at $620$~nm. The arrows indicate the direction of the polarizer. (\textbf{C}) Extinction map of the band of interest in the BZ, obtained by summing 20 iso-frequency contours at wavelengths from $550$ nm to $740$ nm at an interval of $10$~nm. Different colors correspond to different wavelengths, with the colormap shown in \textbf{A}. The polarization-averaged data are shown in the center, with the five dark regions corresponding to the vortices. The polarization-resolved data are shown outside, showing dark strips that rotate around the vortices with the rotation direction and speed corresponding to the sign and magnitude of the topological charges.}
\label{fig3}
\end{figure*}

Take bands 2 and 3 for example, whose dispersion curves are plotted in Fig.~\ref{fig1}A in solid and dashed green lines, respectively. Their mirror eigenvalues on $\Gamma$--M--X--$\Gamma$ can be easily obtained by computing the near-field Bloch wavefunctions at three points -- one point on each of the three mirror-invariant momentum lines. The results~($\pm$ values) are labeled next to their dispersions. For these two bands, the possible charges at $\Gamma$ and M points are $4n-1$ and those at the X point are $2n\pm1$.
Note that bands 2 and 3 are non-degenerate on the three momentum segements excluding the end points~($\Gamma$, M, X), so their mirror eigenvalues are constants on each momentum segment.



The sum of all vortex charges on a 2D dispersion band is zero, equaling to the Euler characteristic~($\chi$) of the 2D BZ torus.
If the charge sum at high-symmetry points~{$\Gamma$, M and X} is non-zero, there must be mobile vortices inside the BZ to neutralize the total charge.
When the system parameter is continuously tuned, the mobile vortices could potentially combine at the high-symmetry points and change their $n$ values. We plot the calculated far-field polarization streamlines in the whole BZ for both bands 2 and 3 in Fig.~\ref{fig1}C to illustrate our arguments. The shaded gray area indicates non-radiative states. The charges within the gray area on band 2 can be inferred to be $-3$ in total, and the charge at M point on band 3 can be inferred to be $-1$. For the following experiments, we focus on the lowest radiative dispersion --- band 2.

To experimentally characterize the band dispersions and the vortices on them, we built a polarization-resolved momentum-space imaging spectroscopy, as illustrated in Fig.~\ref{fig2}A. The working principle is the fact that the information carried by the back focal plane of an objective lens corresponds to the momentum-space~(Fourier-space) information of the radiation field from the sample.
There are two modes of operation. In the first mode, by positioning an imaging spectrometer, whose entrance is conjugate to the back focal plane,
we map out the momentum space one line at a time within the whole visible spectrum; rotating the sample in plane relative to the entrance slit of the imaging spectrometer yields the entire BZ.
In the second mode, we select a wavelength with a band-pass filter (10 nm bandwidth) and image the iso-frequency contour in the entire BZ onto a 2D charged-coupled-device (CCD) camera in one shot; a series of filters yield a full set of iso-frequency contours.
More importantly, using a polarizer, we can determine the far-field polarizations for every momentum state. Therefore, we were able to obtain the dispersion, lifetime, and polarization of nearly all the radiative states in the whole BZ experimentally (details in Methods Summary).

Fig.~\ref{fig2}D shows a 3D plot of the measured band structure of the fabricated square-lattice plasmonic crystal (image in Fig.~\ref{fig2}B). The plotted data are the extinction ratio (one minus reflectivity) as a function of wavelength and wavevector, under unpolarized illumination and averaged over two orthogonal polarizer directions at the output. The band structure is defined by the peaks in the extinction spectra, which are equivalent to absorption maxima since there is no transmission in our system. These extinction~(or absorption) maxima result from the excitations of SPP modes. The measured band structure agrees well with the calculation in Fig.~\ref{fig1}A. The key feature in Fig.~\ref{fig2}D is that the extinction peaks disappear at certain momentum points (marked by the blue arrows). Since dispersion bands are physically continuous, the disappearance indicates that those states cannot be excited and are decoupled from the free space with a diverging radiative quality (Q) factor. Similar states were recently observed in dielectric photonic crystal slabs~\cite{hsu2013observation,gansch2016measurement}. We measured the Q factors of nearly all the Bloch states above the light cone by fitting the experimental extinction spectra with temporal coupled mode theory (detailed in Supplementary Notes). Fig.~\ref{fig2}E shows the distribution of radiative Q factors on band 2 (blue dashed line in Fig. \ref{fig2}D), where a total of five vortices can be seen from the diverging Q. One vortex is fixed at $\Gamma$, and the other four are along the $\Gamma$--X lines. 

Furthermore, we determined the far-field polarizations using the extinction spectra measured at three polarizer angles. As shown in Fig.~\ref{fig2}C, the polarizer angles were $0$, $45$ and $90$ degrees relative to the direction of the entrance slit of the spectrometer. With these information, the polarization direction of the far field of each Bloch state can be determined. The details of the methods are shown in the Supplementary Notes. Fig.~\ref{fig2}F shows the measured polarization distribution in the whole BZ, plotted as streamlines on top of a pseudocolor plot. Five polarization vortices are clearly observed with one $-1$ vortex at $\Gamma$ surrounded by four $+1$ vortices~(related by the $C_{4v}$ symmetry of the lattice). At the core of each vortex, the polarization is ill-defined, indicating an absence of far field and corresponding to the divergent Q factors in Fig.~\ref{fig2}E. Due to their topological nature, polarization vortices are insensitive to the variation of the structure parameters such as filling fractions, which we verified experimentally in the Supplementary Fig. S5. The experimentally obtained topological charges are in a good agreement with the theoretical prediction in Fig.~\ref{fig1}C.

\begin{figure*}[t]
\includegraphics[scale=1.0]{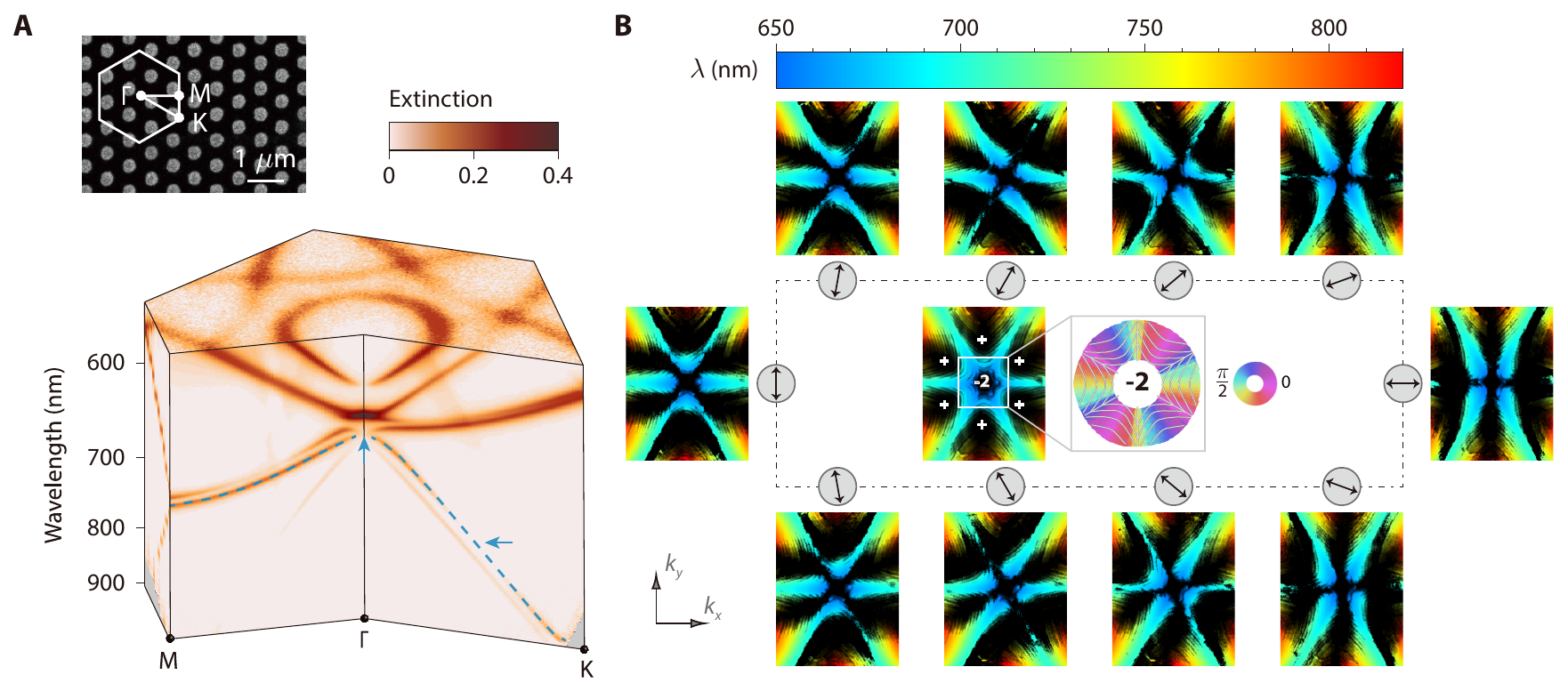}
\caption{\textbf{Observations of high-order polarization vortices in a hexagonal plasmonic lattice.} (\textbf{A}) SEM image of the structure~(upper panel) and the measured 3D band structure using unpolarized illumination~(lower panel). Vortices are marked with blue arrows. (\textbf{B}) Extinction map of the band of interest~(blue dashed line in \textbf{A}), obtained by summing  18 iso-frequency contours at wavelengths from $650$ nm to $820$ nm. The colors correspond to different wavelengths, with the colormap shown at the top. Central plots show the polarization-averaged data and the measured distribution of polarization vectors near the $\Gamma$ point. Outer plots show the polarization-resolved data. The dark strips around $\Gamma$ rotate at half the speed as the polarizer, indicating a vortex with $-2$ topological charge.}
\label{fig4}
\end{figure*}

For an even more direct visualization of the vortices in the BZ, we use the second mode of operation and sum the iso-frequency contours measured at a continuous range of wavelengths. We took a series of polarization-averaged iso-frequency contours from $550$ nm to $740$ nm with a wavelength interval of $10$ nm. The bandwidth of the color filters is also $10$ nm, so the iso-frequency contours overlap, and the whole wavelength range is covered. A few examples of the measured iso-frequency contours are shown in Fig.~\ref{fig3}A and the others are shown in Supplementary Fig. S6. Note that only the colored contours with smaller wavevectors are from band 2, while those in light gray are from the higher bands. By summing the iso-frequency contours contributed by band 2, we obtained the extinction data for the whole dispersion band and plotted it in the central figure in Fig.~\ref{fig3}C. Five dark regions are centered at the vortex cores with near-zero extinction in the momentum space, corresponding to the polarization vortices.

In order to directly resolve the winding of polarization around the vortices, we recorded the iso-frequency contours as a function of the polarizer angle, shown in Fig.~\ref{fig3}B. The outer plots in Fig.~\ref{fig3}C show the summed iso-frequency contours at different polarizer angles; the signals from states with far-field polarization perpendicular to the polarizer diminish, forming a dark strip around each vortex.
Those dark strip patterns spin with the polarizer.
The spinning direction and speed of the dark strips directly reveal the sign and magnitude of the topological charges.
Under the counterclockwise rotation of the polarizer, the spin direction of the vortex at the $\Gamma$ point is clockwise, while other vortices spin counterclockwise. Consequently, they have opposite signs of charges. Since all five vortices rotate at the same speed as that of the polarizer, they have the same amplitude of topological charges of $1$. This direct observation of topological charges is consistent with the numerical results in Fig.~\ref{fig1}C as well as the measured polarization distribution in Fig.~\ref{fig2}F. Among the key experimental results of this work, an animation showing these dark strips spinning with the polarizer around all vortex points is presented in the Supplementary Animations.

The topological charge of a vortex can be an arbitrary signed integer. So far, we have observed $\pm1$ charges in the square lattice plasmonic crystal with $C_{4v}$ point group symmetry. Higher order charges can be found either in the higher bands or in a lattice with a higher symmetry group~\cite{zhen2014topological}. 
Here we took the latter approach and fabricated a hexagonal lattice with $C_{6v}$ symmetry.

As shown in the SEM image in Fig.~\ref{fig4}A, the structure is a $70$-nm PMMA layer patterned with hexagonal lattice on a flat silver substrate (periodicity: $600$ nm; hole diameter: $330$ nm). The measured band structure is shown in Fig.~\ref{fig4}A, while the calculation is shown in the Supplementary Fig. S7. Here, we picked the second band (shown as blue dashed line in Fig.~\ref{fig4}A) to observe vortices with high-order charges. The polarization distribution around the $\Gamma$ point obtained from measured polarization-dependent band dispersions are shown in the central right panel of Fig.~\ref{fig4}B. The topological charge of this central vortex is $-2$.
Away form $\Gamma$, there are six polarization vortices in the $\Gamma$--K directions.
In the polarization-resolved iso-frequency contours (outer plots of Fig.~\ref{fig4}B), the dark strips of these six outside vortices rotate counterclockwise with the same angular speed as that of the polarizer, corresponding to $+1$ charges.
Meanwhile, the one at the $\Gamma$ point rotates clockwise with half the angular speed as that of the polarizer,  corresponding to a vortex with $-2$ charge.
The corresponding animation of these spinning vortices is shown in the Supplementary Animations.

In conclusion, we have experimentally observed momentum-space polarization vortices in the entire BZ of plasmonic crystals.
Vortices with topological charges of $\pm1$ were observed in a square lattice, and topological charges of $-2$ and $+1$ were observed in a hexagonal lattice. Such phenomenon is generic for vector fields~\cite{yang2013topological,mohanta2017emergent} and is an additional topological feature in band structures, not captured by the current topological band theory~\cite{lu2014topological,bansil2016colloquium}.



%

\bigskip

\noindent\textbf{\large Methods Summary}

\noindent\textbf{Sample fabrication.}
The 2D plasmonic crystal studied here is a flat silver film coated with a periodic corrugated thin poly-methylmethacrylate (PMMA) layer fabricated using electron-beam lithography. The silver film was coated on a glass substrate using a thermal evaporator \textsf{(Kurt J. Lesker NANO 36)}. The thickness of the silver film is 200 nm. A thin 70-nm PMMA layer was then spin-coated on the silver film. Desired periodic structures are written on the PMMA layer using electron-beam lithography \textsf{(Zeiss Sigma 300 scanning electron microscopy plus Raith Elphy Plus)}.

\noindent\textbf{Experimental measurement.}
A home-made polarization-resolved momentum-space imaging spectroscopy was built based on an Olympus microscope \textsf{(IX73)}. The incident light is irradiated from a tungsten lamp, focused by an objective \textsf{(40$\times$ magnitude, NA 0.95)} on to a sample. The back focal plane of the same objective was imaged onto the entrance slit of an imaging spectrometer using a series of convex lens. A schematic view of the setup is shown in Fig.~2A. Thus the reflected light from the sample into a given angle corresponds to a single position on the entrance slit. After the spectrometer \textsf{(Princeton Instruments IsoPlane-320)}, the angle-resolved information, in other words the information in momentum space, is imaged with a CCD \textsf{(PIXIS 400)}. A polarizer is placed on the focal plane of lens 1 to select a certain polarization. Normalizing the reflected light from the sample ($I_s$) by that from a silver mirror ($I_m$), we obtain the angle-resolved and polarization-resolved reflection ($R=I_s/I_m$) and extinction (${\rm E}_\mathrm{xt}=1-R$) spectra of the sample. Averaging over two orthogonal polarizer orientations yields unpolarized extinction. High extinctions arise from the excitations of the resonant modes of the structure. In the band structure measurement, one axis of the 2D CCD is used to resolve the angle of the reflected light, while the other axis is used to resolve the wavelength. The band structure $E(k)$ was subsequently calculated from the angle and wavelength resolved spectra as $E=\frac{1240}{\lambda~(\mathrm{nm})}~\mathrm{eV}$ and $k= k_0\sin\theta$, where $k_0=2\pi/\lambda$ and $\theta$ is the angle with respect to sample normal.
Each measurement yields the band structure across one line $\Gamma$--Z$_i$ of the Brillouin zone (as shown in Supplementary Fig.~S4) that is parallel to the entrance slit of the spectrometer. By rotating the sample in plane relative to the entrance slit of the imaging spectrometer, we obtain the band structure inside the grey regions shown in Supplementary Fig.~S4. Due to lattice symmetry, this yields the band structure and polarization information inside the entire Brillouin zone. In the iso-frequency contour measurements, a narrow band-pass filter \textsf{(Delta Optical Thin Film A/S)} was used. The grating of the spectrometer was working on the $0$th order, and the entrance slit was fully opened. Here, both axes of the 2D CCD was used to resolve the angle dependence of the extinction. As mentioned before, there is a one-to-one correspondence between the angle $\theta$ and momentum $k$, thus the iso-frequency contours are obtained. As in the band structure measurements, we also obtain both polarization-resolved and polarization-averaged results. With different pass-bass filters, we obtain a series of iso-frequency contours covering the full range of dispersion of the band of interest.

\bigskip

\noindent\textbf{Acknowledgements} We thank Shaoyu Yin, Chen Fang, Zhong Wang, and Bo Zhen for helpful discussions.
The work was supported by 973 Program (2013CB632701 and 2015CB659400), China National Key Basic Research Program (2016YFA0301100 and 2016YFA0302000) and NSFC (11404064). The research of L. S. was further supported by Science and Technology Commission of Shanghai Municipality (17ZR1442300), Professor of Special Appointment (Eastern Scholar) at Shanghai Institutions of Higher Learning, and the
Recruitment Program of Global Youth Experts (1000 plans).
L.L. was supported by the Ministry of Science and Technology of China under Grant No.~2017YFA0303800, 2016YFA0302400 and the
National Thousand-Young Talents Program of China.

\bigskip

\noindent\textbf{Author contributions} L.L., L.S., and J.Z. conceived the idea of this study. Y.W.Z. fabricated the samples. Y.W.Z and A.C. conducted the measurement. W.Z.L performed theoretical analysis of topological vortices. C.W.H., A.C. and Y.W.Z analysed the data. L.S., L.L. and J.Z. supervised the project. All authors discussed and interpreted the results. L.L., J.Z., L.S. and C.W.H wrote the paper with input from all authors.


\end{document}